\documentstyle[12pt,aaspp4]{article}

\slugcomment{\shortstack[r]{
Submitted to ApJ Letter, 1996 October 4\\
KUNS-1414\\
YITP-96-42}}

\begin{document}

%%%%%%%%%%%%%%%%%%%%%%%%%%%%%%%%%%%%%%%%%%%%%%%%%%%%%%%%%%%%%%%%%%
%%%%%% (1) Title Page   %%%%%%%%%%%%%%%%%%%%%%%%%%%%%%%%%%%%%%%%%%
%%%%%%%%%%%%%%%%%%%%%%%%%%%%%%%%%%%%%%%%%%%%%%%%%%%%%%%%%%%%%%%%%%

\title{Fragmentation of the Primordial Gas Clouds\\ 
and the Lower Limit on the Mass of the First Stars}

\author{Hideya Uehara, Hajime Susa, Ryoichi Nishi, and Masako Yamada}
\affil{Department of Physics, Kyoto University, Kyoto 606-01, Japan}

\and 

\author{Takashi Nakamura}
\affil{Yukawa Institute for Theoretical Physics, Kyoto University, 
Kyoto 606-01, Japan}

%%%%%%%%%%%%%%%%%%%%%%%%%%%%%%%%%%%%%%%%%%%%%%%%%%%%%%%%%%%%%%%%%%
%%%%%% (2) Abstract & Subject Heading  %%%%%%%%%%%%%%%%%%%%%%%%%%%
%%%%%%%%%%%%%%%%%%%%%%%%%%%%%%%%%%%%%%%%%%%%%%%%%%%%%%%%%%%%%%%%%%

\begin{abstract}
We discuss the fragmentation of primordial gas clouds
in the universe after decoupling.
Comparing the time scale of collapse with that of fragmentation,  
we obtain the typical mass of a fragment both numerically and analytically.
It is shown that the estimated mass gives the minimum mass of a fragment which
 is formed from the primordial gas cloud and is essentially determined 
by  the {\it Chandrasekhar mass}.
\end{abstract}

\keywords{galaxies: formation --- molecular processes --- stars: formation}

%%%%%%%%%%%%%%%%%%%%%%%%%%%%%%%%%%%%%%%%%%%%%%%%%%%%%%%%%%%%%%%%%%
%%%%%% (4) Text & Ackowledgment %%%%%% %%%%%%%%%%%%%%%%%%%%%%%%%%%
%%%%%%%%%%%%%%%%%%%%%%%%%%%%%%%%%%%%%%%%%%%%%%%%%%%%%%%%%%%%%%%%%%

\section{Introduction}
The first stars formed after decoupling had no metals
heavier than lithium, so that star formation proceeded without grains. 
This means that  the isothermal approximation, which is often used to 
investigate star formation in the present molecular cloud 
(its validity is explained by Hayashi \& Nakano 
\markcite{HN65}1965), is not applicable and that star formation
in the primordial gas cloud may be completely different.   
Up to now many authors have studied  dynamical and thermal
 evolutions of primordial gas clouds 
(Matsuda, Sato, \& Takeda \markcite{MST69}1969; 
Yoneyama \markcite{Yone72}1972; 
Hutchins \markcite{Hut76}1976;
Yoshii \& Sabano \markcite{YS79}1979; 
Carlberg \markcite{Carl81}1981; 
Palla, Salpeter, \& Stahler \markcite{PSS83}1983; 
Lahav \markcite{Laha86}1986; 
Puy \& Signore \markcite{PD96}1996; 
Susa, Uehara, \& Nishi \markcite{paperI}1996, hereafter paper I).
One of the most important quantity in such  studies is the 
minimum Jeans mass in the primordial gas cloud in relation to 
the initial mass function.
In a spherical  free-falling cloud, however, the density perturbation
grows only in a power law (Hunter 1964)
 as in the expanding universe, so that the
minimum Jeans mass does not necessarily mean the mass of the fragment 
(Larson \markcite{Lars85}1985 and references therein).
In general a collapsing cloud does not necessarily 
fragment at the instant  when the Jeans
 mass reaches the minimum, but fragments when the collapse 
of the cloud is almost halted (see \S \ref{sec2}).
%More quantitatively, a collapsing cloud is likely to fragment when the 
%condition $t_{dyn} \gtrsim t_{frag}$ is satisfied
%(Inutsuka \& Miyama \markcite{IM92}1992 for 
%cylindrical cloud), where $t_{dyn}$ and $t_{frag}$ are the dynamical 
%time scale of the collapse and the growing time scale of the
%fragments, respectively.
\par
It is known that a spherical cloud in pressure-free collapse is unstable 
against 
 non-spherical perturbations (Lin, Mestel, \& Shu \markcite{LMS65}1965; 
Hutchins \markcite{Hut76}1976; paper I). This means that 
the collapsing primordial gas cloud is not spherical even without 
angular momentum in general. Therefore  the evolution of a 
non-spherical cloud must be considered. In one of the plausible
 scenarios,   shock waves occur in the collapsing non-spherical 
primordial gas cloud.
In the post shock flow, due to  cooling by hydrogen molecules
 formed through non-equilibrium recombination, 
a cooled sheet ($T \lesssim 500 \mbox{K}$) is
formed (MacLow \& Shull \markcite{MS86}1986; Shapiro \& Kang 
\markcite{SK87} 1987, hereafter SK).
This cooled sheet  fragments more easily  into filaments  
 (Miyama, Narita, \& Hayashi 
\markcite{MNHa}1987a, \markcite{MNHb}1987b) than into spherical clouds.
As the virial temperature of a filament is essentially 
{\it constant}  (see eq. [\ref{A2}]), 
its evolution is much 
different from that of  a spherical cloud whose  virial temperature 
increases as the  cloud contracts.

In this Letter  we  investigate the dynamical and
 thermal evolution of an infinitely long cylinder to model 
a filament formed in the post shock cooled sheet and 
estimate the minimum mass of  the fragment of the cylinder.
The formulation of our model is shown in \S \ref{sec3}. 
In \S \ref{sec2} we show how to estimate the mass of the
fragment in our model. 
The numerical results are shown in \S \ref{sec4}. We analytically
estimate the final fragment mass in \S \ref{sec5}.  
 \S \ref{sec6} will be devoted to discussions .
\section{Basic equations}
\label{sec3}
\subsection{Equation of motion of the cloud for cylindrical collapse}
\label{s6}
For an isothermal cylinder there is a critical line density $M_c(T)$
(Ostriker \markcite{Ost64}1964)given  by
\begin{equation}
M_c(T)= \frac{2k_{B}T}{\mu m_{\rm H} G} ,            \label{k2}     
\end{equation}
where $ T, \mu $, and $m_{\rm H}$ represent the isothermal
 temperature, the mean molecular weight  and  the mass of
 hydrogen atoms, respectively.
An isothermal cylinder with line density $M$ larger than
 $M_c(T)$  has no equilibrium  and collapses, while  
an isothermal cylinder with $M$ less  
than $M_c(T)$  expands  if the external pressure does not exist.

When the  cylindrical cloud is formed as a result of  the fragmentation
 of the sheet, the gas pressure is comparable to 
the gravitational force. So the effect of the gas pressure must be 
included in the equation of motion of the cylindrical cloud. 
To know the dynamical evolution of such a cylinder we begin with the virial 
equation for a cylinder,
\begin{eqnarray}
\frac{1}{2} \frac{d^2 I}{d t^2}=2 T+2 {\it \Pi} - G M^2 , \label{v1}
\end{eqnarray}
where
\begin{eqnarray}
  I = \int_V \rho r^2 dV   , \;\;\;
  T = \int_V \frac{1}{2} \rho u^2 dV  ,\;\;\;
{\it \Pi} = \int_V p dV   , 
\end{eqnarray}
and the integration is effected over the volume per unit length of 
 the cylinder, $V$.
Here we approximate that the cylinder is uniform and has constant 
density $\rho$, scale radius $R=\sqrt{M/ \pi \rho}$, and  temperature $T$.  
Then above $I, T,\; {\rm and}\; {\it \Pi}$ become
\begin{eqnarray}
  I = \frac{1}{2}M R^2  ,\;\;\;     
  T = \frac{1}{4}M \left ( \frac{dR}{dt} \right )^2  ,\;\;\;
{\it \Pi} = \frac{k_B T}{\mu m_{\rm H}} M \; .  \label{P2}  
\end{eqnarray}
Substituting  (\ref{P2}) in equation (\ref{v1}) we obtain
\begin{eqnarray}
\frac{d^2 R}{dt^2}&=&-\frac{2 G}{R} ( M-M_c(T) ) \; .  \label{eq2}  
\end{eqnarray}
We describe the evolution of the cylinder by this equation.

\subsection{Energy equation and  chemical reactions}
The equations describing thermal processes are the same as those of 
paper I.
The time evolution of the cloud temperature is described by the energy 
equation given by
\begin{equation}
\frac{d \varepsilon}{dt}=-P \frac{d}{dt}\frac{1}{\rho}-\Lambda_{rad} -
\Lambda_{chem}  \;,   \label{z2}
\end{equation} 
where $\varepsilon$, $\Lambda_{rad}$ and $\Lambda_{chem}$   are 
the  thermal energy per unit mass , the cooling by the radiation  and
the cooling/heating by chemical reactions, respectively.
In $\Lambda_{rad}$ we take into account  ${\rm H_2}$ line emissions
due to the vibrational and rotational transitions between excited
levels.
The gas pressure $P$ is evaluated by the equation of state of the ideal gas.
The details of $\Lambda_{rad}$ are shown in Appendix B 
of paper I and $\Lambda_{chem}$ is shown in SK. 
\par
We use the same chemical reaction rates as Palla et al. 
\markcite{PSS83}(1983). 
We consider ${\rm H, H_2, H^+, H^-, \mbox{and}~ e^- }$ as chemical
compositions  of the cloud. 
Helium and its ions are not included in our calculations because they
 are chemically inert at such low temperatures ($T \lesssim 10^4 {\rm K}$) as
 considered here.
Hereafter we use $f_i$ to denote  the fractional number abundance of
 the i-th composition.
\section{The epoch of fragmentation and the mass of a fragment}
\label{sec2}
The dispersion relation of the linear density perturbations of
 an isothermal sheet with surface 
density $\Sigma$ shows  that   the perturbation with wavelength 
 $\lambda_{frag} \sim 4\pi H$  grows 
 fastest, where $H = c_s^2/\pi G \Sigma $ and 
$c_s=\sqrt{k_B T/\mu m_{\rm H}}$ is the sound velocity 
(Simon \markcite{Si65}1965).
Although the primordial gas clouds we consider are not isothermal,  
the dispersion relations are  weakly dependent on  equation of state
(see Fig.1 of Larson \markcite{Lars85}1985) and we use
the isothermal dispersion relations as typical ones. As shown by Miyama
et al. (\markcite{MNHa}1987a, \markcite{MNHb}b), an equilibrium sheet
tends to fragment into filaments, whose line density  is estimated by
\begin{eqnarray} 
 M  =  \lambda_{frag} \Sigma 
   \sim  \frac{4k_B T}{\mu m_{\rm H} G} .   \label{s4}
\end{eqnarray}
Since the above $M$ is greater than $M_c(T)$ 
 the filament formed by fragmentation of an isothermal sheet cannot 
be in an equilibrium configuration but begins to collapse.
As shown by Inutsuka \& Miyama \markcite{IM92}(1992) 
such a cylinder with $M > M_c(T)$ does not  fragment immediately.
To estimate  the epoch of the fragmentation of a collapsing cloud
 we should compare the dynamical time scale of collapse defined by
 $t_{dyn}=\rho/ \frac{d \rho }{dt}$ 
with that of fragmentation, $t_{frag}$.
When $t_{dyn} \lesssim t_{frag} $, the cylinder collapses so fast that 
the density perturbation  does not grow similar to the spherical case. 
The  collapsing cloud fragments when the condition 
\begin{equation}
t_{dyn} \sim t_{frag}    ,     \label{s1}
\end{equation}
is satisfied, i.e. when the collapse almost stops. In such a situation
we can apply the perturbation analysis of the equilibrium sheet and
the cylinder.  
\par
In the case of the isothermal cylinder in equilibrium, the perturbation 
with  wavelength $\lambda_{frag} \sim 2 \pi  R $  grows fastest, 
where $R$ is the scale radius of the cylinder defined by
 $R=\sqrt{M/\pi \rho_0}$  with  $M$ and $\rho_0$ being the 
line density and the central density, respectively. 
The  e-folding time scale of the fragmentation  is given by
$t_{frag} = 2.1 /  \sqrt{2\pi G \rho_0}$,
(Nagasawa \markcite{N87}1987).
The mass of the fragment formed by the fragmentation of the isothermal 
cylinder is estimated by
\begin{eqnarray}
M_{frag}  =  \lambda_{frag} M   
          \sim   2 \pi  R M    .      \label{s5} 
\end{eqnarray}
In our model  we estimate the mass of the fragment by equation  (\ref{s5}) 
at the epoch when the condition (\ref{s1}) is satisfied. 

\par 
\section{The results of numerical calculations}  \label{sec4}
To obtain initial conditions of a cylindrical cloud which is formed
by  the fragmentation of the post shock cooled sheet,
 we recalculate a typical case of  SK 
with  ${\rm z=5}, \Omega_b h^2=0.1$, and shock velocity 
$v_s=200 {\rm km \; s^{-1}}$.
As initial conditions of our cylindrical cloud we adopt physical 
parameters of the post shock one dimensional flow
 at the point when the  cooling time of the sheet
($\sim$ the dynamical time scale of the sheet)  becomes equal 
to the time scale of the fragmentation, so that the sheet is expected
to fragment into filaments.
The initial condition of the cylinder thus obtained is 
$n_0=6.9 \mbox{cm}^{-3}, 
T_0=330 \mbox{K},\:  f_{\rm H_2},_0=2.3 \times 10^{-3}, {\rm and \:}
f_{\rm e^-},_0= 2.7 \times 10^{-4} $. 
Under the above initial condition we calculated for the cases 
$M/M_c(T_0)=1.0,\; 1.3,\; 1.5,\; 1.7,$ and $2.0$ 
(case A$\sim$E, see Table 1),  for comparison.
\par
\begin{figure}
\plotone{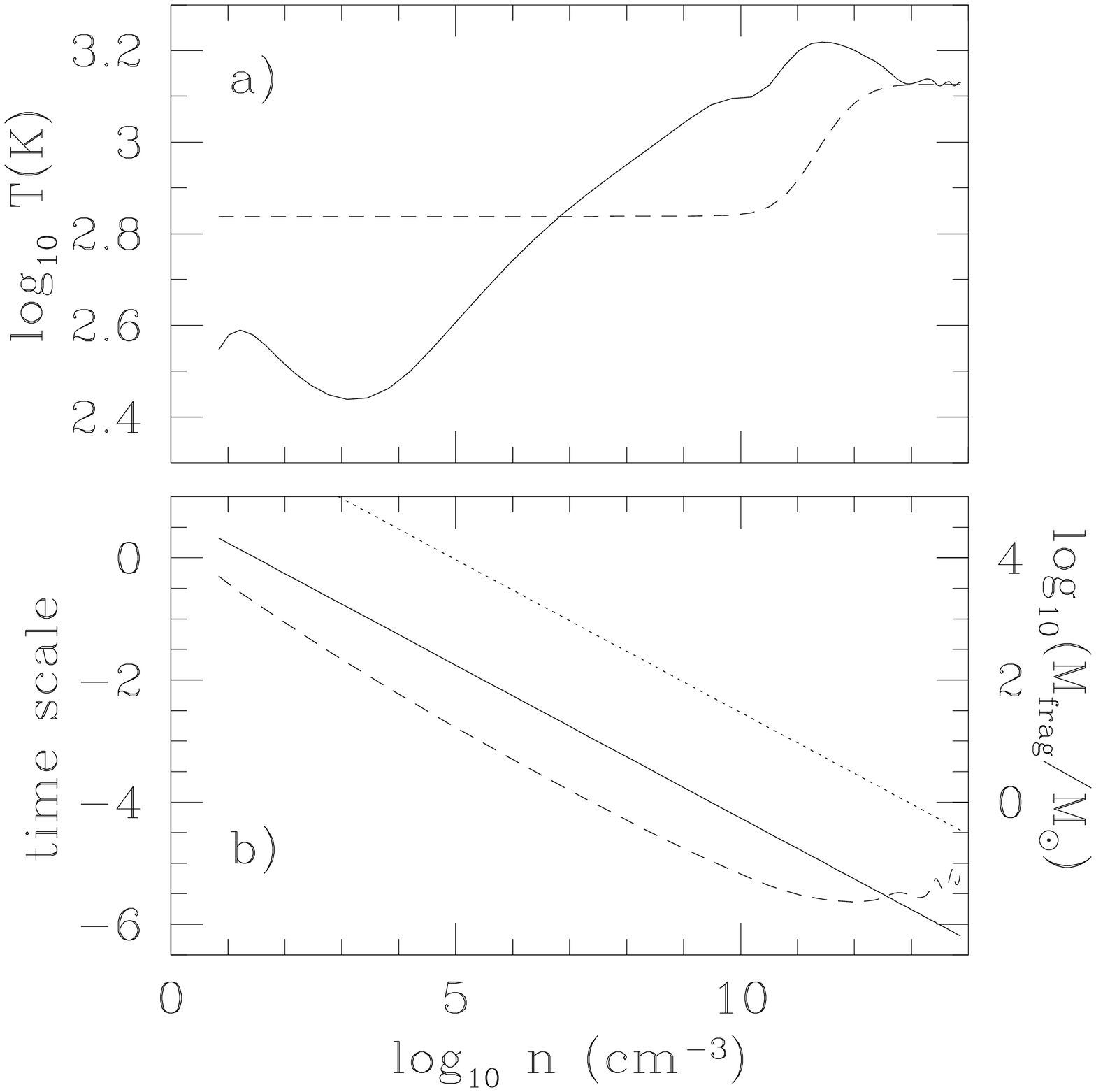}
\caption{a) The thermal evolution of a cylindrical cloud on the n-T plane
(solid curve) for case E. 
The critical temperature is also plotted(dashed curve).
b) Evolution of time scales and mass of a fragment for case E. 
Time scales are normalized by the initial free-fall time.
At first the time scale for collapse, $t_{dyn}$(dashed curve), is
shorter than that for fragmentation, $t_{frag}$(solid curve), by an
order of magnitude. But finally $t_{frag}$ becomes shorter than
 $t_{dyn}$, at which the collapsing cylinder is thought to fragment and
 the mass of a fragment(dotted curve) becomes $\sim 2 M_{\sun}$.
\label{fig1}}
\end{figure}

Since the equilibrium sheet tends to fragment into  the cylinder of line 
density $M \sim 2 M_c(T_0)$ (eq. [\ref{k2}, \ref{s4}]), we show the
evolution for case E as representative, for which 
 the mass of the fragment becomes minimum.
The evolution on the n-T (number density - temperature) 
plane and virial temperature (eq. [\ref{A2}]) are shown in Figure 
\ref{fig1}a. 
 From Figure \ref{fig1}a we can see that the cylinder continues to collapse
even after the cloud becomes  optically thick 
to the line emission of hydrogen molecules 
($n \gtrsim 10^{10} \mbox{cm}^{-3}$). 
 The fact  that the cloud can collapse 
isothermally as a function of time by radiative cooling 
($n \gtrsim 10^{12} \mbox{cm}^{-3}$) is the
characteristic feature of the cylindrical shape of the cloud,
 for which the gravitational force is 
proportional to $R^{-1}$. Since temperature becomes constant, hydrogen
molecules does not dissociate.
Figure \ref{fig1}b shows two time scales $t_{frag}$ and $ t_{dyn}$
 normalized by the initial free-fall time
scale ($1/ \sqrt{2 \pi G \rho_0}$) and  the mass of the fragment 
(eq. [\ref{s5}]).
Comparing two time scales,  we conclude
that the collapsing cylinder fragments  at $n \sim 3 \times 10^{12}
\mbox{cm}^{-3} $ and the mass of the fragment becomes $\sim 2
M_{\sun}$ for case E. 
\par
In Table \ref{table1} number density $n$, hydrogen molecule fraction 
$f_{\rm H_2}$
when the fragmentation occurs (column 3 and  4, respectively), and the mass
of the fragment $M_{frag}$ (column 5)
are listed for each line densities $M$ (column 2). 
From Table \ref{table1} we can see the tendency that the larger the
initial line density $M$, the smaller the mass of the fragment $M_{frag}$ and 
 $M_{frag}$  converges to several solar mass for larger line densities.
Interpretation of these results will be  stated in the next section.

\par
\section{Analytical estimate of the minimum fragment mass}   \label{sec5}
It is interesting to ask what determines the mass of the fragments
 of the collapsing cylinder for case E. 
The collapsing cylinder fragments when the condition (\ref{s1}) is satisfied. 
As the cylinder collapses isothermally as a function of time finally, 
we can set $d \varepsilon/dt \sim 0$ in equation (\ref{z2}) and
from this we obtain $t_{dyn} \sim (\gamma-1) \: t_{cool}$, where
$\gamma=5/2$ for diatomic molecular gas.  
Since  the cloud cools by the line emissions of hydrogen molecules and
is optically thick  to these lines emission,
$t_{cool}$ can be estimated by 
\begin{equation}
t_{cool} \sim  \frac{ \frac{1}{\gamma-1} \frac{M}{\mu m_{\rm H}} k_B T}
              {2\pi R \sigma T^4 \frac{\Delta \nu}{\nu} \alpha_c } ,
\label{A1}
\end{equation}
where $\sigma=2 \pi^5 k_B^4 /15 h^3 c^2 $ is the Stefan-Boltzmann constant, 
 and $\alpha_c$ is the effective number of line emissions.   
Since the optical depth is $\sim 100$ at most, the wing of line profile
may not be
important (Rybicki \& Lightman \markcite{RadiPro}1979) and the line
broadening is determined by Doppler broadening as
${\Delta \nu}/{\nu}={v_{\rm H_2} }/{c}
=\sqrt{ {k_B T}/{m_{\rm H} c^2} } $.
The effects of chemical heating/cooling are not  included in $t_{cool}$ 
because the fraction of hydrogen molecule $f_{\rm H_2}$ is almost  unity 
by the epoch of the fragmentation.  
The temperature  of the cylinder is estimated by the virial
temperature as
\begin{equation}
k_B T=\frac{1}{2}\mu m_{\rm H} G M  . \label{A2}
\end{equation}
 From the above equations we obtain 
\begin{eqnarray}
M_{frag}  \sim  2\pi R M    
 \sim   \sqrt{ \frac{1}{\alpha_c} }
\frac{1}{\mu^{9/4} }\frac{m_{Pl}^3}{m_{\rm H}^2}   ,  \label{A3}
\end{eqnarray}
 where $m_{Pl}=\sqrt{hc/G}$ is the Planck mass.
Equation (\ref{A3}) means that  $M_{frag}$ is essentially equal to
the Chandrasekhar mass ($\sim m_{Pl}^3/m_p^2$, here $m_p$ is proton mass).  
\par
So far we did not consider other heating processes such as
 shocks and turbulence which  likely to occur in the collapsing cloud.
All these processes tend to halt the collapse, so that
the epoch of fragmentation  becomes  earlier and, as a result, the mass of
fragment increases.  
So equation  (\ref{A3}) gives the  lower limit on the mass of the fragment
which is formed from the primordial gas cloud. 
Our estimate shows even the primordial gas cloud which
cools most effectively by hydrogen molecules cannot fragment into
a smaller mass than equation  (\ref{A3}).

\par
For smaller line densities the gas pressure force halts the collapse  
before three-body reactions (Palla et al. \markcite{PSS83}1983)
 {\it completely} convert atomic hydrogens to molecules.
In these cases  equation (\ref{A3}) is not applicable because 
clouds are optically thin to line emissions for 
$f_{\rm H_2} \lesssim 10^{-1}$  and
chemical heating by ${\rm H_2}$ formation must be considered when 
hydrogen molecules are formed  by three-body reactions.
These clouds fragment at lower density, in other words at larger scale radius,
than the density at which  they reach 
the isothermally collapsing stage (corresponding to 
$n \gtrsim 10^{12} {\rm cm}^{-3}$ of case E).
Since  $M_{frag}$ is proportional to the scale radius at fragmentation,
this leads to the larger fragment mass.

\section{Discussions}
\label{sec6}

The opacity-limited hierarchical fragmentation scenario was argued by
several authors 
(Low \& Lynden-Bell \markcite{LL76}1976; 
Rees \markcite{Rees}1976; 
Silk \markcite{Sil76}1977). 
According to Rees \markcite{Rees}(1976), the minimum Jeans mass of 
the spherical cloud is given by 
\begin{eqnarray}
M_F&=&\left( \frac{k_B T}{m_{\rm H} c^2} \right )^{1/4}f^{-1/2}  \nonumber
     \frac{1}{\mu^{9/4} }\frac{m_{Pl}^3}{m_{\rm H}^2}   , \\   \nonumber
\end{eqnarray}
 where f ($\lesssim 1$) is the efficiency of the radiation of the cloud 
compared with the black body radiation  and it depends  on the details of
 the cooling process and the opacity.  
As the present molecular clouds  have various molecules and grains, f is
nearly equal to unity when the cloud becomes optically thick to the
radiation.  
In the case of the primordial gas cloud which is cooled by line emissions
of hydrogen molecules, f is proportional to $\sqrt{T}$ ( Doppler
broadening ) and  $T$-dependence of $M_F$  completely vanishes.  
Apparently this gives the same conclusion as ours.  However previous authors
argued the spherical collapse, which does not lead to fragmentation, 
and our argument would be more relevant.
\par
Although our estimate is order of magnitude and  the exact mass of the
fragment cannot be estimated, 
the existence of the lower limit on the mass of the
fragment we found would have significant meanings.
For example, the following scenario for formation of our Galaxy may be
possible: the existence of lower limit on the fragment mass cause the
mass of the first stars to be  massive, so that the first stars complete
main-sequence stage within the age of the universe and evolve to 
dark remnants (e.g. white dwarfs, neutron stars, and black holes) 
in the Galactic halo after some mass loss, whose gas fall to the
 Galactic disk to form it.
This scenario would be consistent with infall model
(Lynden-Bell \markcite{Ly75}1975; Vader \& de Jong \markcite{VJ81}1981; 
Clayton \markcite{Cl88}1988),
which was proposed to explain G dwarf problem (the observed paucity of
metal-poor stars, see Rocha-Pinto \& Maciel
\markcite{RM96}1996 for recent observation).
Remnants thus formed by the above scenario may be MACHOs, whose estimated
mass ($\sim 0.5 M_{\sun}$, Bennett et al.\markcite{MACHOgr}1996)
is consistent with that of white dwarfs. 
It is very interesting that some of astrophysical problems, including 
non-existence of zero metal stars, may be  explained in a consistent way by
 such a simple scenario. 

\acknowledgments
We would like to thank H. Sato for useful discussions, T. Chiba for
 critical reading, and S. A. Hayward for checking English.
This work is supported in part by Research Fellowships of the Japan Society 
for the Promotion of Science for Young Scientists, No.6894(HU), 3077(HS)
 and by Grant-in-Aid of Scientific Research of the Ministry of Education,
Culture, Science and Sports, No.08740170(RN), No.07640399(TN).

%%%%%%%%%%%%%%%%%%%%%%%%%%%%%%%%%%%%%%%%%%%%%%%%%%%%%%%%%%%%%%%%%%
%%%%%%  (5) Table           %%%%%%%%%%%%%%%%%%%%%%%%%%%%%%%%%%%%%%
%%%%%%%%%%%%%%%%%%%%%%%%%%%%%%%%%%%%%%%%%%%%%%%%%%%%%%%%%%%%%%%%%%

\newpage

\begin{deluxetable}{ccccr}
\tablecaption{initial line density and results of numerical calculations
\label{table1} }
\tablehead{
\colhead{case} &
\colhead{$M/M_c(T_0)$}    & \colhead{$n (\mbox{cm}^{-3})$} & 
\colhead{$f_{\rm H_2}$} & \colhead{$M_{frag}/M_{\sun}$}
}
\startdata
A& 1.0 & $1 \times 10^8$     & $5 \times 10^{-3}$  & $1 \times 10^2$ \nl
B& 1.3 & $1 \times 10^9$     & $4 \times 10^{-2}$  & $4 \times 10^1$ \nl
C& 1.5 & $2 \times 10^{10}$  & $4 \times 10^{-1}$  & $2 \times 10^1$ \nl
D& 1.7 & $5 \times 10^{11}$  & $9 \times 10^{-1}$  & $3 \times 10^0$ \nl
E& 2.0 & $3 \times 10^{12}$  & $1 \times 10^0$     & $2 \times 10^0$ \nl
\enddata
\end{deluxetable}

%%%%%%%%%%%%%%%%%%%%%%%%%%%%%%%%%%%%%%%%%%%%%%%%%%%%%%%%%%%%%%%%%%
%%%%%% (6) References   %%%%%%%%%%%%%%%%%%%%%%%%%%%%%%%%%%%%%%%%%%
%%%%%%%%%%%%%%%%%%%%%%%%%%%%%%%%%%%%%%%%%%%%%%%%%%%%%%%%%%%%%%%%%%

%%%%%%%%%%%%%%%%%%%%%%%%%%%%%%%%%%%%%%%%%%%%%%%%%%%%%%%%%%%%%%%%%%
%%%%%% (7) Figure Captions  %%%%%%%%%%%%%%%%%%%%%%%%%%%%%%%%%%%%%
%%%%%%%%%%%%%%%%%%%%%%%%%%%%%%%%%%%%%%%%%%%%%%%%%%%%%%%%%%%%%%%%%%
\newpage

\newpage

\begin{references}

\reference{AN96}Anninos, P., \& Norman, M. L. 1996, \apj, 460, 556

\reference{MACHOgr}Bennett, D. P., et al. 1996, astro-ph/9606012 preprint

\reference{Carl81}Carlberg, R. G. 1981, \mnras, 197, 1021

\reference{Cl88}   Clayton, D. D. 1988, \mnras, 234, 1

\reference{HN65}Hayashi, C., \& Nakano, T. 1965, Prog. Theor. Phys., 34, 754

\reference{Hun64}Hunter, C.      1964, \apj, 139, 570

\reference{Hut76}Hutchins, J. B. 1976, \apj, 205, 103

\reference{IM92}Inutsuka, S., \& Miyama, S. M. 1992, \apj, 388, 392

\reference{Laha86}Lahav, O.     1986, \mnras, 220, 259

\reference{Lars85}Larson, R. B. 1985, \mnras, 214, 379

\reference{LMS65}Lin, C. C., Mestel, L., \& Shu, F. H. 1965, \apj, 142, 1431  

\reference{LL76}Low, C., \& Lynden-Bell, D. 1976, \mnras, 176, 367

\reference{Ly75}Lynden-Bell, D. 1975, Vistas in Astronomy, 19, 299

\reference{MS86} MacLow, M.-M., \& Shull, J. M.        1986, \apj, 302, 585

\reference{MST69}Matsuda, T., Sato, H., \& Takeda, H. 1969, 
Prog. Theor. Phys., 42, 219

\reference{MNHa}Miyama, S. M., Narita, S., \& Hayashi, C. 1987a, 
Prog. Theor. Phys., 78, 1051

\reference{MNHb}--------.   1987b, Prog. Theor. Phys., 78, 1273 

\reference{N87} Nagasawa, M. 1987, Prog. Theor. Phys., 77, 635

\reference{Ost64}Ostriker, J. 1964, \apj, 140, 1056

\reference{PSS83}Palla, F., Salpeter, E. E., \& Stahler, S. W. 1983, 
\apj, 271, 632

\reference{PD96}Puy, D., \& Signore, M. 1996, \aap, 305, 371

\reference{Rees}Rees, M. J. 1976, \mnras, 176, 483

\reference{RM96}Rocha-Pinto, H. J., \& Maciel, W. J. 1996, \mnras, 279, 447

\reference{RadiPro}Rybicki, G. B., \&  Lightman, A. P. 1979, \\
Radiative Process in Astrophysics (New York:Wiley-Interscience)
\markcite{Rees}

\reference{SK87}Shapiro, P. R., \& Kang, H. 1987, \apj, 318, 32

\reference{Sil76}Silk, J. 1977, \apj, 214, 152

\reference{Si65}Simon, R. 1965, Ann. d'Astrophys., 28, 40

\reference{paperI}Susa, H., Uehara, H., \& Nishi, R. 1996, Prog. Theor. Phys., 
 submitted

\reference{VJ81}Vader, J. P., \& de Jong, T. 1981, \aap, 100, 124

\reference{Yone72}Yoneyama, T. 1972, \pasj, 24, 87

\reference{YS79}Yoshii, Y., \& Sabano, Y. 1979, \pasj, 31, 305

\end{references}
\end{document}